\documentclass[prb,twocolumn,showpacs]{revtex4}

\usepackage{graphicx}
\usepackage{epsfig}

\makeatletter

\begin{document}

\title{Electronic and Structural Properties of a 4$d$-Perovskite: \\
 Cubic Phase of SrZrO$_3$}
\author{E. Mete, R. Shaltaf, and \c{S}.~Ellialt{\i}o\u{g}lu\footnote{Corresponding author. E-mail: sinasi@metu.edu.tr}}

\affiliation{Department of Physics, Middle East Technical University, Ankara
06531, Turkey}

\date{\today}

\begin{abstract}

First-principles density functional calculations are performed
within the local density approximation to study the electronic
properties of SrZrO$_3$, an insulating 4$d$-perovskite, in its
high-temperature cubic phase, above 1400 K, as well as the generic
3$d$-perovskite SrTiO$_3$, which is also a $d^0$-insulator and
cubic above 105 K, for comparison reasons. The energy bands,
density of states and charge density distributions are obtained
and a detailed comparison between their band structures is
presented. The results are discussed also in terms of the existing
data in the literature for both oxides.

\end{abstract}

\pacs{71.15.Mb, 71.20.-b}

\keywords{ab-initio pseudopotential method, transition metal oxides,
\emph{d}-band perovskites, energy bands, density of states}

\maketitle

\section{Introduction}\label{section1}

The class of transition metal oxides constitutes a big family of interesting
materials with extra physical properties due to the additional d-electrons they
posses. They come in variety of crystal structures and exhibit individually
several of these phases. They include insulators, metals, semiconductors and
also superconductors. Some have delocalized d-bands providing catalytically
active surfaces, some have narrow d-bands with emphasized electron correlations
giving rise to diverse properties like high temperature superconductivity, and
colossal magnetoresistance. They are well known with their ferroelectric,
antiferroelectric and piezoelectric properties. Their use in technological
application is also diverse, including optical wave guides, laser-host
crystals, high temperature oxygen sensors, surface acoustic wave devices,
non-volatile memories, dynamic random access memories, frequency doublers,
piezoelectric actuator materials, and high-K capacitors in various
applications.

Strontium titanate, SrTiO$_3$, is a generic representative of transition metal
oxides which have perovskite crystalline structure. It has been extensively
studied both theoretically and experimentally because of its several
interesting physical and technological properties. It is highly insulating at
room temperature, and in the form of n-type thin films it shows
superconductivity at low temperatures. It is a cubic perovskite at room
temperature with a tetragonal phase transition at 105 K. Its surfaces are very
flat and stable both mechanically and chemically which makes it best electrode
in photocatalysis of water~\cite{erdman}, and makes it best buffer layer for
the growth of gallium arsenide on silicon~\cite{droopard}, and makes it best
substrate for the growth of high T$_c$ cuprate superconductors~\cite{aruta}.
Due to its high dielectric constant it is also one of the leading candidates to
replace the silica as a gate material in silicon technology.

Another insulating perovskite is the strontium zirconate, SrZrO$_3$, with
4$d$-electrons, which is of interest because of its high temperature electronic
properties. Large single crystals of SrZrO$_3$ with high perfection can be
grown with recent techniques and this enables their usage as laser-host
materials and as substrate materials. It was also suggested by Shende {\sl et
al.}~\cite{shende} that these materials can be used in high-voltage capacitor
applications because of their high breakdown strengths as well as high
dielectric constant. In addition to this, both SrTiO$_3$ and SrZrO$_3$ are
suitable for use in high-temperature applications such as fuel cells, steam
electrolysis and hydrogen gas sensors~\cite{iwahara1,kurita,yajima}. This is
because when these type of transition metal oxides are doped with acceptor ions
they exhibit protonic conduction at high temperatures~\cite{iwahara2}.

Unlike SrTiO$_3$, at room temperature SrZrO$_3$ has an
orthorhombic phase as revealed by the first structural studies
which date back to 1960's~\cite{roth,swanson}. Later the existence
of two additional phases at high-temperature was proposed by
Carlsson to be both tetragonal~\cite{carlsson}, however, more
recent studies~\cite{LignyRichet,brendan,matsuda} on high
temperatures have shown that SrZrO$_3$ undergoes three structural
phase transitions summarized as follows: First, orthorhombic
($Pnma$) to orthorhombic ($Cmcm$) at 970 K, then to tetragonal
($I4/mcm$) at 1100 K, and then to cubic ($Pm3m$) at 1400 K. This
compound has a rather high melting temperature of about 2920
K~\cite{souptel}, consequently it is cubic in a wide range of
temperature where most of its useful applications take place.

In this work we have made first-principles pseudopotential calculations of the
electronic band structure, density of states and charge densities for SrZrO$_3$
in the cubic perovskite phase. In addition to that we have made a reference
calculation for SrTiO$_3$ in order to give a discussion by comparison. We have
also made some comparisons with the related experimental data where available.

\section{Calculation Method}
\label{section2}

We used pseudopotential method based on density functional theory
in the local density approximation (LDA). The self consistent norm
conserving pseudopotentials are generated by using the
Troullier-Martins scheme~\cite{TM} which is included in the
fhi98PP package~\cite{fhi98pp}. Plane waves are used as a basis
set for the electronic wave functions. In order to solve the
Kohn-Sham equations~\cite{kohn-sham}, conjugate gradients
minimization method~\cite{Payne} is employed as implemented by the
ABINIT code~\cite{abinit}. The exchange-correlation effects are
taken into account within the Perdew-Wang scheme~\cite{PerdewWang}
as parameterized by Ceperly and Alder~\cite{CeperlyAlder}.

Pseudopotentials are generated using the following electronic
configurations: For Sr 5$s$ electrons are considered as the true
valence. Moreover the 4$s$ and 4$p$ semicore states are added to
the valence states. For O only the true valence states (2$s$ and
2$p$) are taken into account, because these states are enough to
have the correct transferability property. For Ti 4$s$ and 3$d$
true valence states plus the 3$s$ and 3$p$ semicore states are
treated as valence states. Similarly, for the same group element
Zr, 5$s$ and 4$d$ true states and additionally the 4$s$ and 4$p$
semicore states are considered as valence states. Inclusion of
semicore states in the case of Sr, Ti and Zr is required in order
to get the correct electronic properties of these elements in
various physical systems. In other words, the inclusion of these
semicore states make the corresponding pseudopotentials closer to
the all-electron potentials. The above configuration is found to
be the optimized choice for these materials.

All of the calculations involve 5-atom cubic unit cell arranged in a perovskite
structure. We get a good convergence for the bulk total energy calculation with
the choice of cut-off energies at 30 Ha for SrTiO$_3$ and at 33 Ha for
SrZrO$_3$ using $4\times 4\times 4$ Monkhorst-Pack~\cite{MonkhorstPack} mesh
grid. We have found that in the band structure calculations 76 {\bf k}-points
are enough to obtain good results for both of these transition metal oxides. In
the density of states calculations, however, the irreducible Brillouin zone was
sampled with 560 and 455 {\bf k}-points for SrTiO$_3$ and SrZrO$_3$,
respectively.

\section{Results and Discussion}
\label{section3}

The results for structural parameter calculations are summarized
in Table \ref{table1}. The calculated lattice parameters for
SrTiO$_3$ and SrZrO$_3$ are both within 0.5\% of the experimental
results. Likewise the calculated bulk moduli for SrTiO$_3$ and
SrZrO$_3$ are found to be 0.3\% and 0.8\% smaller than the
effective experimental values, respectively. The agreement with
the experiments can be considered to be very good. In the case of
SrTiO$_3$, comparisons of our results with the theoretical work of
Kimura {\sl et al.}~\cite{kimura} and with the calculated values
of van Benthem {\sl et al.}~\cite{vanBenthem} suggest that our
pseudopotentials are as reliable and perform slightly better. To
our knowledge no first-principles calculation is available for
SrZrO$_3$ to compare with.

For the bulk modulus of SrZrO$_3$, we have listed in Table
\ref{table1} an extrapolated value of 150 GPa by Ligny and
Richet~\cite{LignyRichet}, however, such extrapolation is known to
give about 15\% underestimation for the {\bf Zr}-compound in a
series of Ca{\bf M}O$_3$~\cite{Ross}, and by the same token we
expect the experimental value to be higher than 150 GPa.

\begin{table}[htb]
\begin{tabular}{c|cc|cc}\hline\hline
$\rule[2mm]{0mm}{2mm}$ & \multicolumn{2}{c|}{\hskip2mm Lattice Parameter
({\AA})} & \multicolumn{2}{c}{\hskip2mm
Bulk Modulus (GPa)} \\ & \hskip6mm Calc. & Exp. & \hskip6mm Calc. & Exp. \\
\hline $\rule[2mm]{0mm}{2mm}$ \hskip3mm SrTiO$_3$ \hskip3mm & \hskip6mm 3.878 &
\hskip3mm 3.905 ~\cite{MitsuiNomura} \hskip3mm & \hskip6mm 191 & \hskip3mm 183
~\cite{MitsuiNomura} \hskip3mm \\ \hskip3mm SrZrO$_3$ \hskip3mm & \hskip6mm
4.095 & \hskip3mm 4.109 ~\cite{SmithWelch} \hskip3mm & \hskip6mm 171 &
\hskip3mm 150 ~\cite{LignyRichet} \hskip3mm \\ \hline\hline
\end{tabular}
\vskip3mm\caption{Calculated and experimental values for lattice parameter and bulk
modulus of SrTiO$_3$ and SrZrO$_3$\label{table1}}
\end{table}

\begin{figure*}[ht]
\vspace{5mm} \epsfig{file=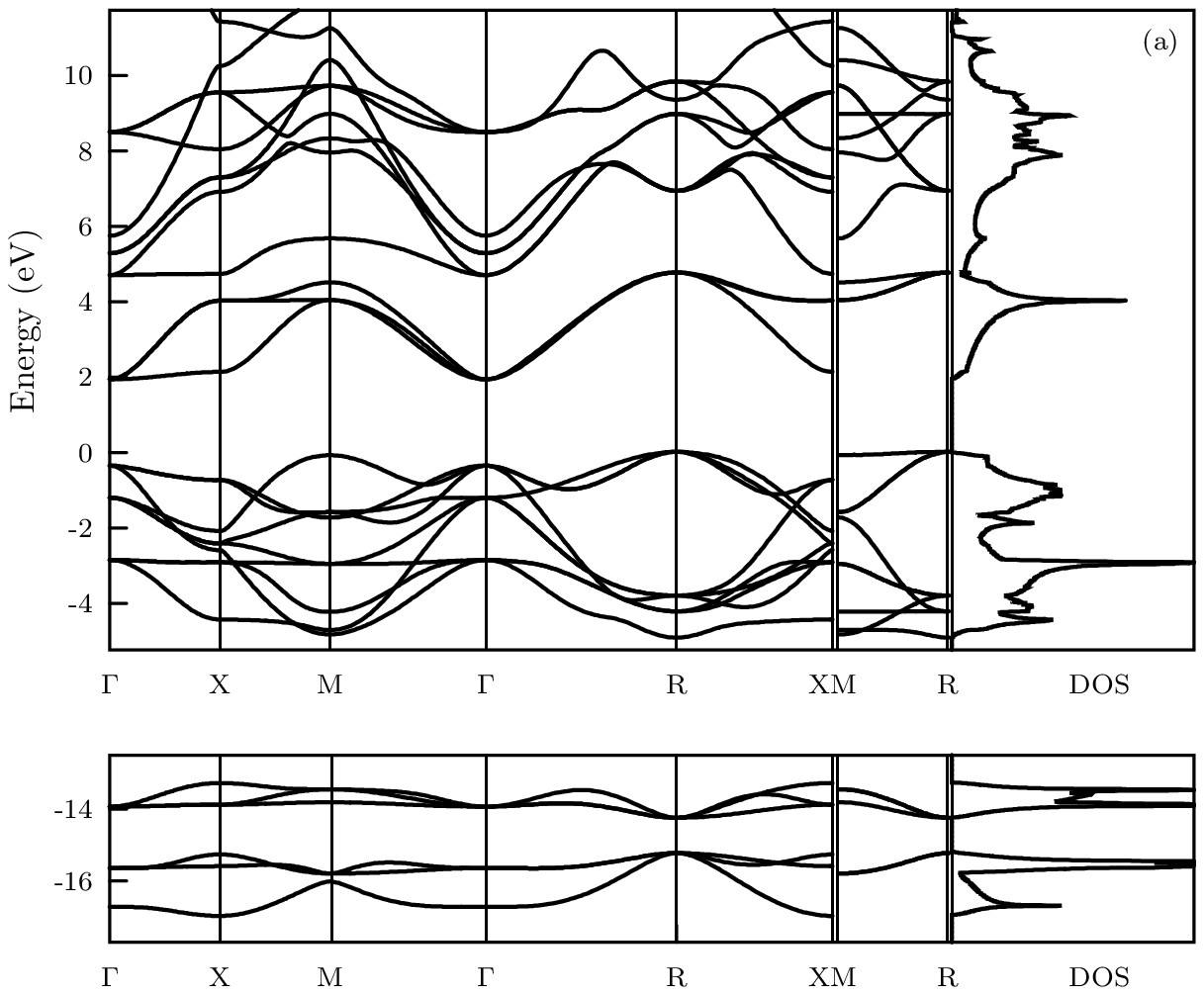,width=8.3cm,height=8.7cm} \hskip8mm
\epsfig{file=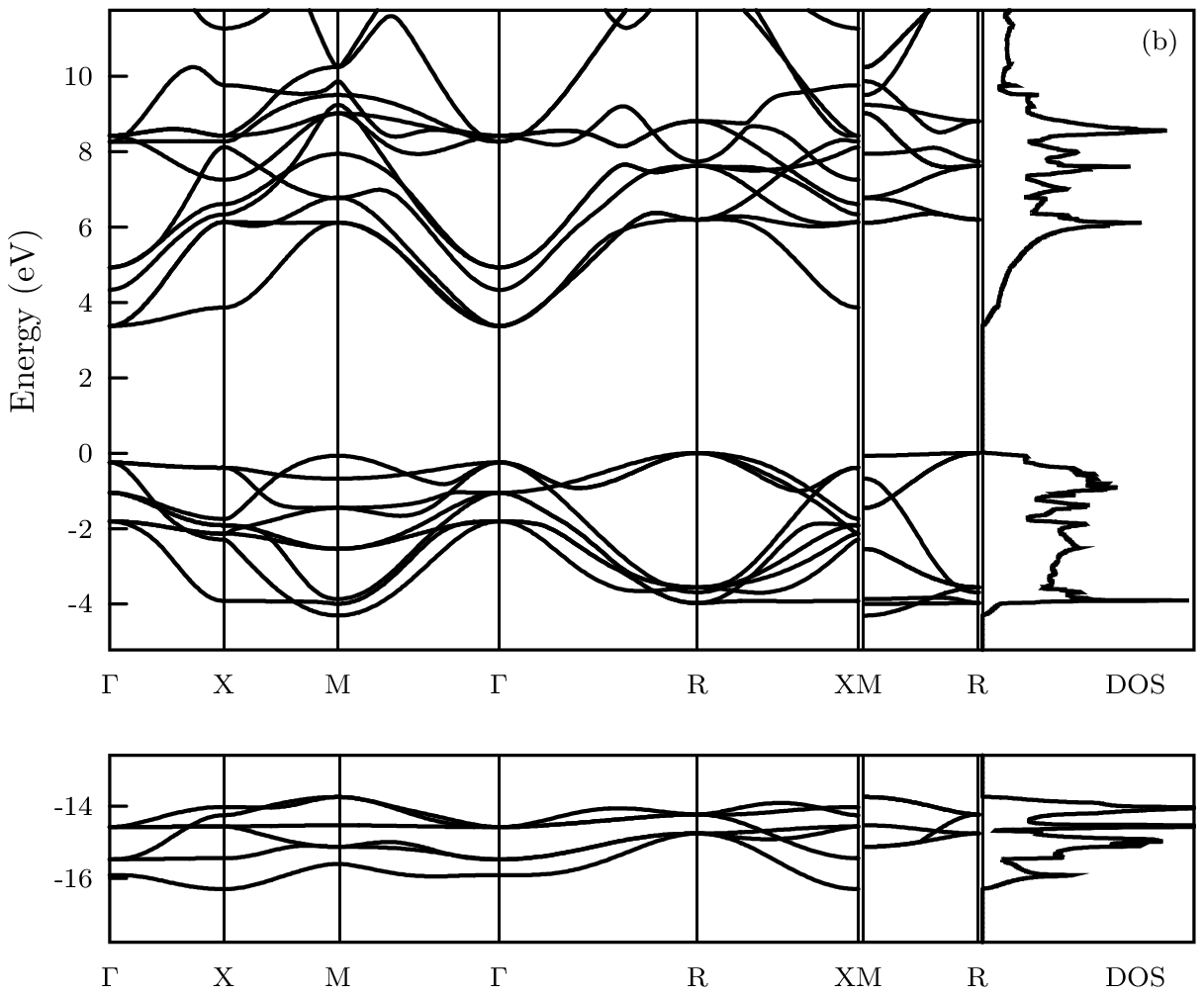,width=8.3cm,height=8.7cm} \caption{Calculated band
structures for (a) SrTiO$_3$ and (b) SrZrO$_3$}\label{figure1}
\end{figure*}

Energy band structures and densities of states are given in Fig.~\ref{figure1},
where the zero of energy is chosen to coincide with the top of the valence
band. The general features of the energy bands are similar for both oxides. An
overall look at the two band structures shows that the lower valence bands are
composed of O 2$s$ and Sr 4$p$ semicore states grouped together at about --15
eV. Although the individual bands have similar bandwidths for both materials,
those of SrTiO$_3$ do not overlap and separated by a $\sim$1 eV gap, whereas
those of SrZrO$_3$ having stronger interaction with each other, especially
around X point, causing an overlap of the corresponding densities of states,
and consequently, the combined bandwidth is smaller.

The upper valence bands have the same trend, i.e., wider for SrTiO$_3$ (4.93
eV) and narrower for SrZrO$_3$ (4.32 eV). Top of the valence bands reflect the
$p$ electronic character mostly due to oxygen-oxygen interaction (down to about
--4 eV) for both of the transition metal oxides. This agrees well with the
DV-X$\alpha$ molecular orbital study of Yoshino {\sl et al.}~\cite{yoshino}.
Even though whole valence band structure is dominated by O 2$p$ states, there
are, however, a mixture of $\sigma$-bands stemming from the $pd\sigma$
interactions extending throughout the whole valence band width, and $\pi$-bands
due to $pd\pi$ interactions which are narrower than that.

Both of SrTiO$_3$ and SrZrO$_3$ have their valence band maxima at the R point.
The energy values of the uppermost band at $\Gamma$ and M points lie slightly
lower than its value at the point R. In SrTiO$_3$ bottom of the valence band
occurs at R, whereas in SrZrO$_3$ it occurs at M point. In both cases, however,
these energy eigenvalues are close to each other at M and R points with a
difference, larger one being for the SrZrO$_3$, of only 0.25 eV.

The lower conduction bands are mainly Ti 3$d$ or Zr 4$d$ states hybridized with
some O 2$p$ electrons giving rise to antibonding $\pi^*$- and $\sigma^*$-bands.
In the case of SrTiO$_3$ the conduction band ordering is as follows: Ti 3$d$
t$_{2g}$ ($\pi^*$-triplet) bands stand alone next to the gap with no overlap to
the Ti 3$d$ e$_g$ states ($\sigma^*$-doublet) which lie just above, and upper
parts of which are mixed with the lower extension of Sr 5$s$ and 4$d$ t$_{2g}$
bands. This ordering is slightly different for SrZrO$_3$, especially at high
energies. Zr 4$d$ t$_{2g}$ states are next to the gap and their upper parts are
mixed with Sr 5$s$ and 4$d$ t$_{2g}$ states. Zr 4$d$ e$_g$ states lie further
up in the energy region between 8.1 eV and 12.5 eV. Conduction band minimum
occurs at $\Gamma$ point in both materials. The lowest conduction band along
$\Gamma$X is more dispersed in strontium zirconate, and consequently the
conduction band edge of its density of states is of three dimensional nature.
Calculated $\pi^*$-conduction band widths are 2.83 eV for SrTiO$_3$ and 3.39 eV
for SrZrO$_3$.

For SrTiO$_3$ the calculated indirect energy band gap between
$\Gamma$ and R points is found to be 1.92 eV and the direct band
gap is 2.30 eV. Corresponding experimental values are 3.25 eV and
3.75 eV~\cite{vanBenthem}, respectively. The results obtained for
SrZrO$_3$ are on the other hand as follows: Indirect band gap is
3.37 eV between $\Gamma$ and R points. The direct band gap being
3.62 eV, is again smaller than the experimental value of 5.9
eV~\cite{lee}. This disagreement between the calculations and
experimental values resulting in narrower theoretical band gaps is
a well known artifact~\cite{JonesGunnarson} of LDA and does not
have any significant effect on the rest of the band structure.

Looking at the density of states pictures one observes several structures
common to both materials. Most of these correspond to singularities in the
bands. There are three flat bands in the conduction band of both materials.
Lowest one is Ti (Zr) t$_{2g}$ band along XM at about 4 eV (6 eV) which causes
the well defined $\pi^*$ peak with a logarithmic van Hove singularity in the
density of states characteristic of two dimensionality of these bands. The next
one up is the Ti (Zr) e$_g$ band along $\Gamma$X at 4.68 eV (8.24 eV) causing a
jump discontinuity in the density of states. Similar $\sigma^*$ shoulder in
density of states is caused by the third flat band which is again Ti (Zr) e$_g$
band along MR at about 9 eV (12.5 eV, not shown).

Valence bands are slightly different in terms of van Hove singularities.
SrTiO$_3$ has a very significant flat band at --2.87 eV along
$\Gamma$XM$\Gamma$ which is a non-bonding $\sigma_0$ band due to O 2$p$ and
responsible for the highest peak at the center of the valence band. Ti 3$d$
t$_{2g}$ band at the same energy along $\Gamma$X is rather flat, and at about
--4.5 eV along XM and also RX it is quite flat close to the X-side. In
addition, the band at --4.25 eV along MR looks almost flat but the
corresponding density of states shows three dimensional behavior just slightly.
For the SrZrO$_3$ valence band, at about --4 eV, the Zr t$_{2g}$ state along
XM, MR and RX are the only flat bands causing the highest peak located at the
bottom of the valence band. The band corresponding to the oxygen non-bonding
state located at the center of SrTiO$_3$ valence band density of states is
dispersed in all directions for SrZrO$_3$.

\begin{figure*}[t]
\begin{tabular*}{\hsize}[t]{cc}
\hskip12mm \epsfig{file=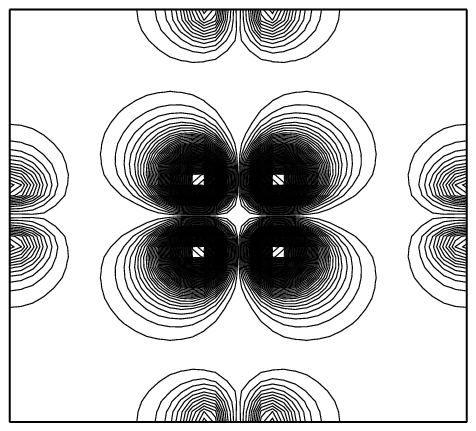,width=5.0cm,clip=} & \hskip15mm
\epsfig{file=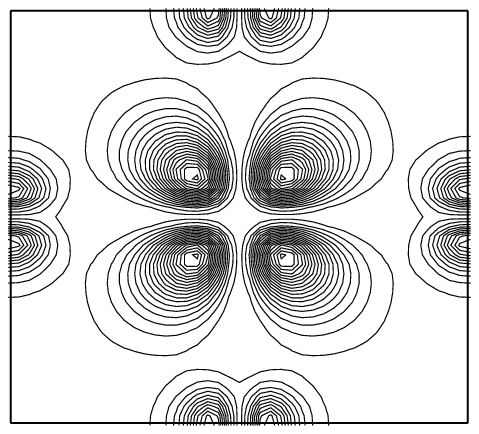,width=5.0cm,clip=} \\
\hskip12mm \epsfig{file=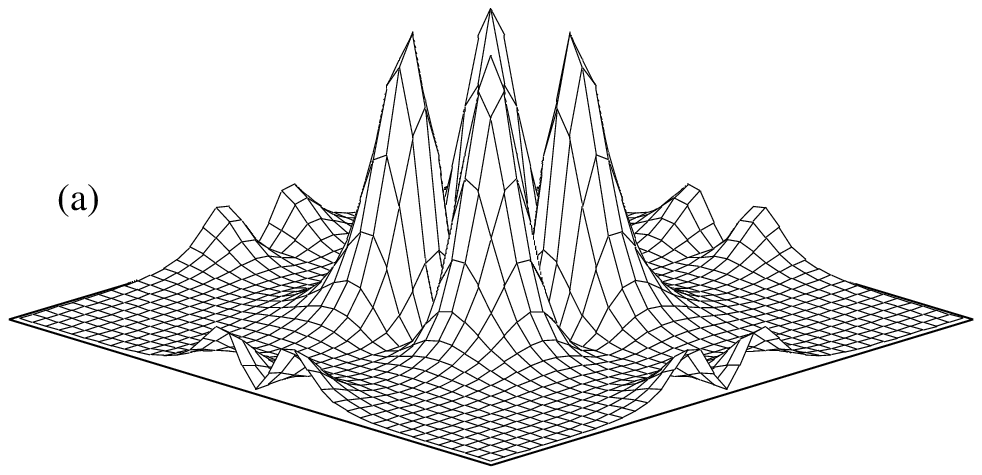,width=7.0cm,clip=} & \hskip15mm
\epsfig{file=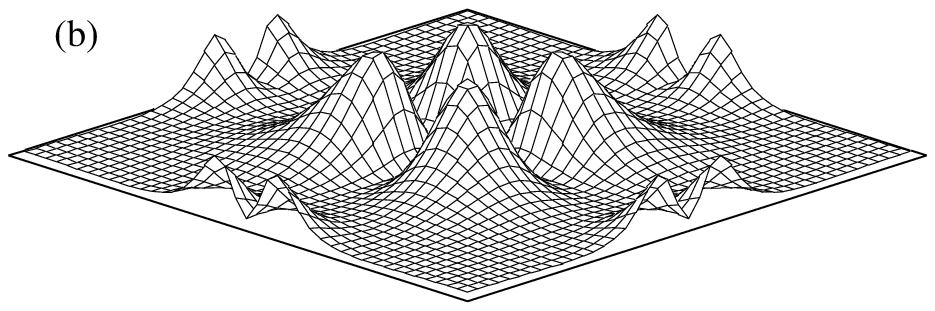,width=7.0cm,clip=} \\
\end{tabular*}
\caption{Charge density contour plots for $\pi^*$ bands for (a)
SrTiO$_3$ and (b) SrZrO$_3$}\label{figure2}
\end{figure*}

Charge density plots were obtained for the lowest three conduction bands
degenerate at the $\Gamma$ point for a (001) plane containing the transition
metal and the four neighboring oxygens. Fig.~\ref{figure2} shows the
composition of these bands to be clearly the hybridization between transition
metal $d$-orbitals with t$_{2g}$ symmetry and O 2$p$ orbitals. The
corresponding $\pi^*$ bands shown in Fig.~\ref{figure1} are more singled out
for SrTiO$_3$ and consequently the charge is more localized as compared to
SrZrO$_3$ whose $\pi^*$ bands are not separated from the rest of the conduction
bands.

\section{Conclusion}
\label{section4}

The structural and electronic properties of two $d^0$-insulator metal oxides,
SrZrO$_3$ and SrTiO$_3$, with cubic perovskite structure are studied using an
ab-initio pseudopotential method. Structural parameters are found to compare
well with the available data in the literature. A detailed description of their
energy bands are given. Corresponding density of states are presented and the
major structures in them are identified. Charge density functions are displayed
for the lower conduction bands for both oxides. Our results for the electronic
properties of SrTiO$_3$ are shown to agree with other calculations and
experimental findings whereas those of SrZrO$_3$ are compared only with the
estimations of Lee {\sl et al.}~\cite{lee} from their optical conductivity
spectra, since to our knowledge, no theoretical calculations exist in the
literature.

\section{Acknowledgement}

This work was supported by T{\"U}B\.{I}TAK, The Scientific and Technical
Research Council of Turkey, Grants No. TBAG-2036 (101T058).

\end{document}